\documentclass{article}
\usepackage{fullpage}
\usepackage{graphicx}
\usepackage{amsmath}
\usepackage{amssymb}
\title{The Massive Photon Impact Factor with Exact Gluon Kinematics}
\date{}
\renewcommand{\vec}[1]{\mbox{\boldmath$ #1 $}}

\begin{document}
\bibliographystyle{utphys}
\newcommand{\msbar}{\ensuremath{\overline{\text{MS}}}}
\newcommand{\DIS}{\ensuremath{\text{DIS}}}
\newcommand{\abar}{\ensuremath{\bar{\alpha}_S}}
\newcommand{\bb}{\ensuremath{\bar{\beta}_0}}
\setlength{\parindent}{0pt}

\titlepage
\begin{flushright}
Cavendish-HEP-2006/14 \\
SPhT-T06/060\\
\end{flushright}

\vspace*{0.5cm}

\begin{center}
{\Large \bf The Virtual Photon-Gluon Impact Factor with Massive Quarks and Exact Gluon Kinematics}

\vspace*{1cm}
\textsc{C.D. White$^{a,}$\footnote{cdw24@hep.phy.cam.ac.uk}, R. Peschanski$^{b,}$\footnote{pesch@spht.saclay.cea.fr}and R.S. Thorne$^{c,}$\footnote{thorne@hep.ucl.ac.uk}} \\

\vspace*{0.5cm} $^a$ Cavendish Laboratory, University of Cambridge, \\ J J Thomson Avenue,
Cambridge, CB3 0HE, UK

\vspace*{0.5cm} $^b$ Service de Physique Th\'{e}orique, CEA/Saclay, \\ 91191 Gif-sur-Yvette cedex, France\\

\vspace*{0.5cm} $^c$  Department of Physics \& Astronomy, University College London, \\ Gower Street, London, WC1E 6BT, UK
\end{center}

\vspace*{0.5cm}

\begin{abstract}
We calculate the impact factor coupling a virtual photon to a gluon via a massive quark-antiquark pair at LL order, but with the imposition of the correct gluon kinematics. Exact analytical results are presented in triple Mellin space with respect to scaled Bjorken $x$, gluon transverse momentum and heavy quark mass. The application of these results to the calculation of approximate NLL coefficient functions needed to relate structure functions to the BFKL gluon is presented. The NLL effects with running coupling are seen to lead to a suppression of the small $x$ divergence when compared with the fixed and running coupling LL results, but less than in the massless case.
\end{abstract}

\vspace*{0.5cm}

\section{Introduction}
\label{intro}
Two factors affecting the successful description of structure function data in deep inelastic scattering are the production of heavy quark flavours in the final state, and high energy corrections to the perturbation expansion due to the BFKL equation \cite{BFKL}. A recent global fit \cite{TW1} with LL order small $x$ resummations including running coupling effects has shown evidence that resummations can improve the fit to current small $x$ data, but also that one should include higher orders in both the fixed order and resummed QCD expansions. For a complete NLL analysis one needs the corresponding impact factors coupling an incoming virtual photon to the BFKL gluon via a quark pair. Even the massless results are at present unknown, although work is in progress \cite{Bartels,FadinImpact}. The massive results will certainly not be available in the near future. However, it is possible to estimate the missing NLL information by supplementing the LL results with information describing the correct kinematic behaviour of the gluon. The LL massless impact factors with exact gluon kinematics were calculated in \cite{Peschanski} and in \cite{TW2} were shown to approximate well the missing NLL information. The purpose of this letter is to extend the calculation of \cite{Peschanski} to the massive quark case. This is of phenomenological importance as it allows an estimate of NLL effects in the heavy gluon coefficient functions which relate the BFKL gluon to measured structure functions. Furthermore, the result has a bearing on the interpretation of the dipole model with heavy quarks beyond LL order (see \cite{Peschanski2} for a discussion). 
\section{Calculation of the Impact Factors}
The presentation closely follows that of the massless case presented in \cite{Peschanski}. Hence we list only the main steps here and omit some details, referring the reader to the previous paper. The total virtual photon-gluon cross-section is given in \cite{Martin}. Taking care to separate the transverse and longitudinal contributions correctly, one has:
\begin{equation}
\sigma_{L,T}\equiv\frac{4\pi^2\alpha}{Q^2}F_{L,T}=\int\frac{dN}{2\pi\imath}(x')^{-N}\int\frac{d\gamma}{2\pi\imath}\tilde{g}_N(\gamma)S_{L,T}(N,\gamma,Q^2/M^2)(Q^2)^{\gamma-1},
\label{sigmal}
\end{equation}
where $\tilde{g}(\gamma)$ is a double Mellin transform of the gluon density, given by:
\begin{equation}
g(x_g,k^2)=\int\frac{dN}{2\pi\imath}x_g^{-N}\int\frac{d\gamma}{2\pi\imath}(k^2)^\gamma\tilde{g}_N(\gamma),
\label{gluon}
\end{equation}
and $\sigma_{L,T}$ is a function of the scaled Bjorken $x$ variable:
\begin{equation}
x'=x\left(1+\frac{4M^2}{Q^2}\right)
\label{x'}
\end{equation}
with kinematic limits $0\leq x'\leq1$, which takes into account the fact that the photon-gluon centre of mass energy must exceed the threshold for heavy quark pair production. Hence, we take moments with respect to $x'$ rather than $x$.  
The variable $x_g$ appearing in equation (\ref{gluon}) is the longitudinal momentum fraction of the gluon that couples to the photon, at the top of the BFKL ladder. This is not the same as the Bjorken $x$ variable, which corresponds to the bottom of the ladder. One may relate them using \cite{Martin}:
\begin{equation}
x_g=x\frac{[(\vec{p}-(1-z)\vec{k})^2+\hat{Q}^2+\hat{k}^2+M^2]}{\hat{Q}^2};\quad \hat{Q^2}=z(1-z)Q^2,
\label{exactkin}
\end{equation}
From these definitions we obtain 
\begin{align}
S_L(N,\gamma,Q^2/M^2)&=4\alpha\alpha_S\left(1+\frac{4M^2}{Q^2}\right)^N\int_0^1dz[z(1-z)]^2\int dk^2\left(\frac{k^2}{Q^2}\right)^{(\gamma-2)}\notag\\
&\int d^2p\left(\frac{1}{p^2+\hat{Q}^2+M^2}-\frac{1}{(\vec{p}-\vec{k})^2+\hat{Q^2}+M^2}\right)^2K_N;
\label{SLint}
\end{align}
\begin{align}
S_T(N,\gamma,Q^2/M^2)&=\alpha\alpha_S\left(1+\frac{4M^2}{Q^2}\right)^N\int\frac{dk^2}{k^2}\int_0^1dz\left(\frac{k^2}{Q^2}\right)^{\gamma-1}\int d^2p\Bigg\{[z^2+(1-z)^2]\left(\frac{\vec{p}}{\vec{p}^2+\hat{Q}^2+M^2}\right.\notag\\
&\left.-\frac{(\vec{p}-\vec{k})}{(\vec{p}-\vec{k})^2+\hat{Q}^2+M^2}\right)^2+M^2\left(\frac{1}{\vec{p}^2+\hat{Q}^2+M^2}-\frac{1}{(\vec{p}-\vec{k})^2+\hat{Q}^2+M^2}\right)^2\Bigg\}K_N.
\label{STint}
\end{align}
The kinematic factor
\begin{equation}
K_N=\frac{(\hat{Q}^2)^N}{[(\vec{p}-(1-z)\vec{k})^2+\hat{Q}^2+\hat{k}^2+M^2]^N}
\label{KN}
\end{equation}
in equations (\ref{SLint},\ref{STint}) is the same as in \cite{Peschanski} up to the 
simple addition of $M^2$ in the denominator. 
This factor came about in \cite{Peschanski} due to the relationship between 
$x$ and $x_g$ and the fact that the Mellin transformation of the cross-section is 
taken with respect to the former while that of the gluon is taken with respect to the latter. 
However, for the case of massive quarks we have the additional 
factor of  
\begin{equation}
\biggl(1+\frac{4M^2}{Q^2}\biggr)^N
\label{x0}
\end{equation}
due to the use of $x'$ rather than $x$ as the real conjugate variable to $N$. Hence, we have an overall effective kinematic factor
\begin{equation}
K^H_N=\frac{(\hat{Q}^2+4\hat{M}^2)^N}{[(\vec{p}-(1-z)\vec{k})^2+\hat{Q}^2+\hat{k}^2+M^2]^N}
\label{KHN}
\end{equation}
representing the difference  between the cross-section variable $x'$ and the gluon momentum 
fraction $x_g$.  \\

One may rewrite equation (\ref{SLint}) as:
\begin{equation}
S_L(N,\gamma,Q^2/M^2)=8\alpha\alpha_S(Q^2)^{2-\gamma}\left(1+\frac{4M^2}{Q^2}\right)^N(A-B),
\label{SLint2}
\end{equation}
where:
\begin{equation}
A=\int_0^1dz[z(1-z)]^2\int\frac{dk^2}{k^4}(k^2)^\gamma\int d^2p\frac{1}{(p^2+\hat{Q}^2+M^2)^2}K_N;%\frac{(\hat{Q})^2}{[(\vec{p}-(1-z)\vec{k})^2+\hat{Q^2}^N+\vec{k}^2+M^2]^N};
\label{A}
\end{equation}
\begin{equation}
B=\int_0^1dz[z(1-z)]^2\int\frac{dk^2}{k^4}(k^2)^\gamma\int d^2p\frac{1}{(p^2+\hat{Q^2}+M^2)[(\vec{p}-\vec{k})^2+\hat{Q}^2+M^2]}K_N.%\frac{(\hat{Q}^2)^N}{[(\vec{p}-(1-z)\vec{k})^2+\hat{Q}^2+\vec{k}^2+M^2]^N}.
\label{B}
\end{equation}
The integrals over $\vec{k}$ and $\vec{p}$ can be evaluated using the procedure of \cite{Peschanski}. The results are:
\begin{align}
A&=\frac{2\pi\Gamma(\gamma-1)\Gamma(N-\gamma+1)}{\Gamma(N)(N-\gamma+2)}(Q^2)^{\gamma-2}\int_0^1dzz^{N+3-\gamma}(1-z)^{N+2}\notag\\
&\times\left[z(1-z)+\frac{M^2}{Q^2}\right]^{\gamma-N+2} \phantom{!}_2F_1(\gamma-1,N-\gamma+1;N-\gamma+3;z);\label{A2}\\
B&=\frac{\pi\Gamma(\gamma-1)\Gamma(N-\gamma+2)}{N\Gamma(N)(N-\gamma+2)}(Q^2)^{\gamma-2}\int_0^1 dz z^{N-\gamma+4}(1-z)^{N+2}\notag\\
&\times\left[z(1-z)+\frac{M^2}{Q^2}\right]^{\gamma-N-2}\phantom{!}_2F_1(\gamma-1,N-\gamma+2;N-\gamma+3;z),\label{B2}
\end{align}
where $\phantom{!}_2F_1(a,b;c;z)$ is the hypergeometric function. It does not seem possible to perform analytically the remaining integration over the (anti-)quark momentum fraction $z$ as in the massless case. \\%However, explicit results are possible after a further Mellin transform in $M^2$, as discussed in the following section.\\

Similar formulae are obtainable for the transverse impact factor. First write:
\begin{equation}
S_T=2\alpha\alpha_S[(DT)+(DT)'-(NDT)-(NDT)'](Q^2)^{1-\gamma}\left(1+\frac{4M^2}{Q^2}\right)^N,
\label{ST2}
\end{equation}
where:
\begin{align}
(DT)&=\int_0^1dzA_T(z)\int dk^2(k^2)^{\gamma-2}\int d^2p\frac{\vec{p}^2}{(\vec{p}^2+\hat{Q}^2+M^2)^2}K_N;\label{DT}\\
(NDT)'&=\int_0^1dzA_T(z)\int dk^2(k^2)^{\gamma-2}\int d^2p\frac{\vec{p}\cdot(\vec{p}-\vec{k})}{(\vec{p}^2+\hat{Q}^2+M^2)[(\vec{p}-\vec{k})^2+\hat{Q}^2+M^2]}K_N;\label{NDT}\\
(DT)'&=M^2\int_0^1dz\int dk^2(k^2)^{\gamma-2}\int d^2p\frac{1}{(\vec{p}^2+\hat{Q}^2+M^2)}K_N\label{DT'};\\
(NDT)'&=M^2\int_0^1dz\int dk^2(k^2)^{\gamma-2}\int d^2p\frac{1}{(\vec{p}^2+\hat{Q}^2+M^2)[(\vec{p}-\vec{k})^2+\hat{Q}^2+M^2]}K_N\label{NDT'}
\end{align}
and $A_T(z)=z^2+(1-z)^2$. Now introduce also the integrals:
\begin{align}
I&=\int\frac{dk^2}{k^4}(k^2)^\gamma\int d^2p\frac{\hat{Q}^2+M^2}{(\vec{p}^2+\hat{Q}^2+M^2)^2}K_N\label{I};\\
J&=\int\frac{dk^2}{k^4}(k^2)^\gamma\int d^2p\frac{\hat{Q}^2+M^2}{[\vec{p}^2+\hat{Q}^2+M^2][(\vec{p}-\vec{k})^2+\hat{Q}^2+M^2]}K_N\label{J};\\
L&=\int\frac{dk^2}{k^4}(k^2)^\gamma\int d^2p\frac{\vec{k}^2}{(\vec{p}^2+\hat{Q}^2+M^2)[(\vec{p}-\vec{k})^2+\hat{Q}^2+M^2]}K_N,\label{L}\\
\end{align}
so that:
\begin{equation}
(DT)-(NDT)=\int_0^1dzA_T(z)\left[J(z)-I(z)+\frac{1}{2}L(z)\right].
\label{JIL}
\end{equation}
Each of the integrals in equations (\ref{DT'}-\ref{L}) can be calculated by analogy with the longitudinal case. The results are:
\begin{align}
(DT)'&=(Q^2)^{\gamma-1}\frac{M^2}{Q^2}\frac{\pi\Gamma(\gamma-1)\Gamma(N-\gamma+1)}{\Gamma(N)(N-\gamma+2)}\int_0^1\left[z(1-z)+\frac{M^2}{Q^2}\right]^{\gamma-N-2}\notag\\
&\quad\times z^{N-\gamma+1}(1-z)^N\phantom{!}_2F_1(\gamma-1,N-\gamma+1;N-\gamma+3;z)\label{DT'int}\\
(NDT)'&=(Q^2)^{\gamma-1}\frac{M^2}{Q^2}\frac{2\pi\Gamma(\gamma-1)\Gamma(N-\gamma+2)}{\Gamma(N+1)(N-\gamma+2)}\int_0^1dz\left[z(1-z)+\frac{M^2}{Q^2}\right]^{\gamma-N-2}\notag\\
&\quad\times z^{N-\gamma+2}(1-z)^N\phantom{!}_2F_1(\gamma-1,N-\gamma+2;N-\gamma+3;z)\label{NDT'int}\\
\int_0^1dz A_T(z)I(z)&=(Q^2)^{\gamma-1}\frac{\pi\Gamma(\gamma-1)\Gamma(N-\gamma+1)}{\Gamma(N)(N-\gamma+2)}\int_0^1dz\left[z(1-z)+\frac{M^2}{Q^2}\right]^{\gamma-N-1}\notag\\
&\quad\times z^{N-\gamma+1}(1-z)^N[1-2z(1-z)]\phantom{!}_2F_1(\gamma-1,N-\gamma+1;N-\gamma+3;z);\label{Iint}\\
\int_0^1dz A_T(z)J(z)&=(Q^2)^{\gamma-1}\frac{2\pi\Gamma(\gamma-1)\Gamma(N-\gamma+2)}{\Gamma(N+1)(N-\gamma+2)}\int_0^1dz\left[z(1-z)+\frac{M^2}{Q^2}\right]^{\gamma-N-1}\notag\\
&\quad\times z^{N-\gamma+2}(1-z)^N[1-2z(1-z)]\phantom{!}_2F_1(\gamma-1,N-\gamma+2;N-\gamma+3;z);\label{Jint}\\
\int_0^1dz A_T(z)L(z)&=(Q^2)^{\gamma-1}\frac{2\pi\Gamma(\gamma)\Gamma(N-\gamma+1)}{\Gamma(N+1)(N-\gamma+1)}\int_0^1dz\left[z(1-z)+\frac{M^2}{Q^2}\right]^{\gamma-N-1}\notag\\
&\quad\times z^{N-\gamma+1}(1-z)^N[1-2z(1-z)]\phantom{!}_2F_1(\gamma,N-\gamma+1;N-\gamma+2;z).\label{Lint}
\end{align}
Putting things together, the total impact factors may be written as:
\begin{align}
S_L(N,\gamma,Q^2/M^2)&=\frac{8\pi\alpha\alpha_S\Gamma(\gamma-1)\Gamma(N-\gamma+1)}{\Gamma(N)(N-\gamma+2)}\left(1+\frac{4M^2}{Q^2}\right)^N\int_0^1\left[z(1-z)+\frac{M^2}{Q^2}\right]^{\gamma-N-2}z^{N-\gamma+3}(1-z)^{N+2}\notag\\
&\times\left[\phantom{!}_2F_1(\gamma-1,N-\gamma+1;N-\gamma+3;z)-\frac{2(N-\gamma+1)z}{N}\phantom{!}_2F_1(\gamma-1,N-\gamma+2;N-\gamma+3;z)\right]\label{SLinttot}\\
S_T(N,\gamma,Q^2/M^2)&=\frac{2\pi\alpha\alpha_S\Gamma(\gamma-1)\Gamma(N-\gamma+1)}{\Gamma(N)(N-\gamma+2)}\left(1+\frac{4M^2}{Q^2}\right)^N\int_0^1\left[z(1-z)+\frac{M^2}{Q^2}\right]^{\gamma-N-1}z^{N-\gamma+1}(1-z)^N\notag\\
&\times \left\{z(1-z)\left[2-\left(z(1-z)+\frac{M^2}{Q^2}\right)^{-1}\right]\Big[\phantom{!}_2F_1(\gamma-1,N-\gamma+1;N-\gamma+3;z)\right.\notag\\
&-\frac{2(N-\gamma+1)z}{N}\phantom{!}_2F_1(\gamma-1,N-\gamma+2;N-\gamma+3;z)\Big]+\frac{(\gamma-1)(N-\gamma+2)}{N(N-\gamma+1)}\notag\\
&\times[1-2z(1-z)]\phantom{!}_2F_1(\gamma,N-\gamma+1;N-\gamma+2;z)\Bigg\}\label{STinttot}
\end{align}
We also note that one can use equation (\ref{SLint}) to rederive the known LL impact factor \cite{Catani2} in a simpler form. Setting $N=0$ in this equation, one may carry out the integrals over $k^2$ and $\vec{p}$ as in the massless case presented in \cite{Forshaw} to give:
\begin{equation}
S_L(0,\gamma,Q^2/M^2)=4\pi\alpha\alpha_S\frac{\Gamma^3(1-\gamma)\Gamma(\gamma)}{\Gamma(2-2\gamma)}\frac{(1-\gamma)}{(3-2\gamma)}\int_0^1dz[z(1-z)]^2\left[z(1-z)+\frac{M^2}{Q^2}\right]^{\gamma-2}.
\label{SLintLL}
\end{equation}
The integral over $z$ can be performed after making the change of variable:
\begin{displaymath}
z=\left\{\begin{array}{c}(1-\sqrt{1-u})/2,\quad z<1/2\\(1+\sqrt{1-u})/2,\quad z>1/2,\end{array}\right.
\end{displaymath}
followed by $u'=1-u$ to get:
\begin{equation}
S_L(0,\gamma,Q^2/M^2)=\frac{2\pi\alpha\alpha_S}{15}\frac{\Gamma^3(1-\gamma)\Gamma(\gamma)}{\Gamma(2-2\gamma)}\frac{(1-\gamma)}{(3-2\gamma)}\left(\frac{M^2}{Q^2}\right)^{\gamma-2}\left(1+\frac{Q^2}{4M^2}\right)^{\gamma-2}\phantom{!}_2F_1\left(2-\gamma,\frac{1}{2};\frac{7}{2};\frac{Q^2}{Q^2+4M^2}\right).
\label{SLintLL2}
\end{equation}
This can be shown to be equivalent to the previously published result for the massive LL longitudinal impact factor\footnote{Note that a factor of $\gamma/(4\pi^2\alpha)$ is needed to convert to the notation of the impact factor $h_L(\gamma,M^2/Q^2)$ as defined by Catani, Ciafaloni and Hautmann.}\cite{Catani2}, which has a slightly more complicated analytic form instead of being proportional to a single hypergeometric term. No particular simplification is possible for the transverse impact factor, however, as this has more terms in the integral definition (\ref{STint}).
\section{Results in Triple Mellin Space}
\label{tripmel}
It seems that the integrals in the preceding section cannot be performed analytically due to the factors of $[z(1-z)+M^2/Q^2]$ in the integrand. However, one may decouple the mass dependence and obtain closed analytic forms for the impact factors by taking a further Mellin transform with respect to $M^2/Q^2$, defined by:
\begin{equation}
S_i(N,\gamma,M^2/Q^2)=\left(1+\frac{4M^2}{Q^2}\right)^N\int\frac{d\gamma_1}{2\pi\imath}\left(\frac{M^2}{Q^2}\right)^{1-\gamma_1}\tilde{S}_i(\gamma,N,\gamma_1),
\label{meldef}
\end{equation}
where the kinematic factor of $(1+4M^2/Q^2)$ from $x'$ has also been included. Using the result:
\begin{equation}
\int dy\, y^{\gamma_1-2}[z(1-z)+y]^b=\frac{\Gamma(\gamma_1-1)\Gamma(1-\gamma_1-b)}{\Gamma(-b)}[z(1-z)]^{b+\gamma_1-1},
\label{melresult}
\end{equation}
one finds from equation (\ref{SLinttot}):
\begin{align}
\tilde{S}_L(N,\gamma,\gamma_1)&=\frac{8\pi\alpha\alpha_S\Gamma(\gamma-1)\Gamma(\gamma_1-1)\Gamma(N-\gamma+1)\Gamma(N+3-\gamma_1-\gamma)}{\Gamma(N)\Gamma(N-\gamma+3)}\int_0^1dz\,z^{\gamma_1}(1-z)^{\gamma+\gamma_1-1}\notag\\
&\times\left[\phantom{!}_2F_1(\gamma-1,N-\gamma+1;N-\gamma+3;z)-\frac{2(N-\gamma+1)z}{N}\phantom{!}_2F_1(\gamma-1,N-\gamma+2;N-\gamma+3;z)\right].\label{tildeS1}
\end{align}
The integral over $z$ now reduces to a conventional hypergeometric integral, and one may use the standard result \cite{Stegun}:
\begin{equation}
\int_0^1dx\,x^{\alpha-1}(1-x)^{\beta-1}\phantom{!}_2F_1(a,b;c;x)=\frac{\Gamma(\alpha)\Gamma(\beta)}{\Gamma(\alpha+\beta)}\phantom{!}_3F_2(a,b,\alpha;c,\alpha+\beta;1)
\label{inthyper}
\end{equation}
to yield:
\begin{align}
\tilde{S}_L(N,\gamma,\gamma_1)&=\frac{8\pi\alpha\alpha_S\Gamma(N+3-\gamma_1-\gamma)\Gamma(\gamma_1-1)\Gamma(\gamma+\gamma_1)}{\Gamma(N)}\notag\\
&\times \Bigg[\phantom{!}_3G_2(\gamma-1,N-\gamma+1,\gamma_1+1;N-\gamma+3,\gamma+2\gamma_1+1;1)\notag\\
&-\frac{2}{N}\phantom{!}_3G_2(\gamma-1,N-\gamma+2,\gamma_1+2;N-\gamma+3,\gamma+2\gamma_1+2;1)\Bigg],
\label{SLtilde2}
\end{align}
where we have used the Meijer G function $\phantom{!}_3G_2(a,b,c;d,e;z)=\Gamma(a)\Gamma(b)\Gamma(c)/[\Gamma(d)\Gamma(e)]\phantom{!}_3F_2(a,b,c;d,e;z)$. One can check this result against the massless result for the impact factor after the integration over $z$ has been performed \cite{Peschanski}. The limit $M^2\rightarrow 0$ corresponds to $\gamma_1\rightarrow 1$ according to the Mellin variable definition of equation (\ref{meldef}) and so one has:
\begin{align}
\lim_{\frac{M^2}{Q^2}\rightarrow0}\int_0^\infty\frac{dM^2}{Q^2}\left(\frac{M^2}{Q^2}\right)^{\gamma_1-2}S_L(N,\gamma,M^2/Q^2)&=\tilde{S}_L(N,\gamma,\gamma_1)|_{\gamma_1\rightarrow 1}\notag\\
&=\lim_{\frac{M^2}{Q^2}\rightarrow0}\int_0^\infty\frac{dM^2}{Q^2}\left(\frac{M^2}{Q^2}\right)^{\gamma_1-2}S_L(N,\gamma,0)\notag\\
&=\left[\Gamma(\gamma_1-1)S_L(N,\gamma,0)\right]_{\gamma_1\rightarrow 1}.\notag
\end{align}
Thus equation (\ref{tildeS1}), divided by $\Gamma(\gamma_1-1)$, reduces as $\gamma_1\rightarrow 1$ to the corresponding result in the massless case [5]. When $M=0$ a simplification occurs, and one can eliminate the Meijer G functions in favour of $\psi$ functions. Such a simplification is not possible in the massive case of equation (\ref{SLtilde2}) because of the non-zero $\gamma_1$. \\

Similarly, one can Mellin transform equation (\ref{STinttot}) to $\gamma_1$-space, with the result:
\begin{align}
\tilde{S}_T(N,\gamma,\gamma_1)&=\frac{2\pi\alpha\alpha_S\Gamma(2-\gamma_1-\gamma+N)\Gamma(\gamma_1-1)\Gamma(\gamma_1+\gamma)(N-\gamma+1)}{\Gamma(N)}\notag\\
&\times\Bigg\{2\phantom{!}_3G_2(\gamma-1,N-\gamma+1,\gamma_1+1;N-\gamma+3,\gamma+2\gamma_1+1;1)\notag\\
&-\frac{4}{N}\phantom{!}_3G_2(\gamma-1,N-\gamma+2,\gamma_1+2;N-\gamma+3,\gamma+2\gamma_1+2;1)\notag\\
&+\frac{1}{N(N-\gamma+1)}\Big[\frac{1}{\gamma_1+\gamma-1}\phantom{!}_3G_2(\gamma,N-\gamma+1,\gamma_1;N-\gamma+2,\gamma+2\gamma_1-1;1)\notag\\
&-2\phantom{!}_3G_2(\gamma,N-\gamma+1,\gamma_1+1;N-\gamma+2,\gamma+2\gamma_1+1;1)\Big]\notag\\
&+\frac{(2-\gamma_1-\gamma+N)}{(\gamma_1+\gamma-1)(N-\gamma+1)}\Big[\frac{2}{N}\phantom{!}_3G_2(\gamma-1,N-\gamma+2,\gamma_1+1;\gamma+2\gamma_1,N-\gamma+3;1)\notag\\
&-\phantom{!}_3G_2(\gamma-1,N-\gamma+1,\gamma_1;\gamma+2\gamma_1-1,N-\gamma+3;1)\Big]\Bigg\}.\label{STtilde2}
\end{align}  
\section{Implications for Structure Functions}
The impact factors discussed in this paper contain some of the information in the unknown NLL heavy impact factors. In the massless case, we have found evidence that the exact kinematics impact factors approximate well the true NLL calculation \cite{TW2}. The comparison relied upon knowledge of the fixed order coefficient and splitting functions $P_{qg}$ and $C_{Lg}$ up to NNLO \cite{Vogtcl,Vogt_s}. A similar analysis is not possible for the heavy impact factors, as the NNLO heavy coefficient and splitting functions are not known, and thus there is no subleading small $x$ information at fixed order available for comparison with resummed results. Nevertheless, by analogy with the massless case it is a reasonable assumption that the NLL impact factors are approximated well by the exact kinematics results and we can therefore use them to investigate the relevant phenomenological quantities. First one interprets the impact factors in terms of the coefficient functions relating the BFKL gluon density to the proton structure functions using the $k_T$ factorisation formula \cite{Collins,CCH} in double Mellin space:
\begin{equation}
F_i^H=C_{i,g}(\gamma,N,M^2/Q^2){\cal G}(\gamma,N),
\end{equation}
where $F_i^H$ ($i\in\{2,L\}$) is the heavy flavour contribution to the structure function\footnote{Note that this contribution corresponds to a {\it fixed flavour} description of the structure function, where the number of active quark flavours is the same at all $Q^2$. In practice one should use a {\it variable flavour scheme} at high $Q^2$ but we do not consider such a scheme here.}. Here ${\cal G}(\gamma,N)$ is the unintegrated gluon density, related to the integrated gluon in Mellin space by ${\cal G}(\gamma,N)=\gamma g(\gamma,N)$. Taking into account also the normalisation factor in equation (\ref{sigmal}), one identifies:
\begin{align}
C_{L,g}(\gamma,N,M^2/Q^2)&=\frac{\gamma}{4\pi^2\alpha} S_L(\gamma,N,M^2/Q^2);\label{CLg}\\
C_{2,g}(\gamma,N,M^2/Q^2)&=\frac{\gamma}{4\pi^2\alpha} [S_T(\gamma,N,M^2/Q^2)+S_{L}(\gamma,N,M^2/Q^2)].\label{C2g}
\end{align}
Expressions for these quantities in the physical space of $x'$ and $Q^2$ are obtained by solving the BFKL equation. We here adopt the solution method of \cite{Thorne01} to investigate the behaviour of the NLL resummed coefficient functions. The calculation demonstrates that the results of equations (\ref{SLinttot}, \ref{STinttot}) are of phenomenological use even though the final integration in $z$ cannot be performed analytically.\\

For a NLL solution of the BFKL equation, one requires the impact factors truncated at ${\cal O}(N)$ and expanded as a power series in $\gamma$. Expansion of the integrands of equations (\ref{SLinttot}, \ref{STinttot}) is possible using the following integral representation of the hypergeometric function \cite{Stegun}:
\begin{equation}
\phantom{!}_2F_1(a,b;c;z)=\frac{\Gamma(c)}{\Gamma(b)\Gamma(c-b)}\int_0^1dv\,v^{b-1}(1-v)^{c-b-1}(1-vz)^{-a}.
\label{intrep}
\end{equation}
Then one has, for example:
\begin{align}
&\phantom{!}_2F_1(\gamma-1,N-\gamma+1;N-\gamma+3;z)=\frac{\Gamma(N-\gamma+3)}{\Gamma(N-\gamma+1)}\int_0^1dv\,(1-v)(1-vz)\bigg\{1+N\log{v}+{\cal O}(N^2)\notag\\
&-\gamma\big[\log(1-vz)+\log{v}+\big(\log{v}\log(1-vz)+\log^2{v}\big)N+{\cal O}(N^2)\big]+{\cal O}(\gamma^2)\bigg\}.
\label{hypexpand}
\end{align}
This is a power series in $N$ and $\gamma$, of the form:
\begin{equation}
\phantom{!}_2F_1(\gamma-1,N-\gamma+1;N-\gamma+3;z)=\sum_{n}\sum_{m}N^n\gamma^m\int_0^1dv\,f_{nm}(v,z),
\label{hypexpand2}
\end{equation}
where the integrands $f_{nm}(t,z)$ become progressively more complicated as $n$ and $m$ increase. A few low order terms can be integrated analytically. When this is not possible, one can parameterise the integral using a suitable function of $z$:
\begin{equation}
\int_0^1dv\,f_{nm}(v,z)\simeq \sum_r k_{nm}^{(1)r} z^r+\sum_s k_{nm}^{(2)s} (1-z)\log^s(1-z)
\label{fitfunc}
\end{equation}
for some range of $r$ and $s$. Values of the integral can be taken for various values of $z$, and fitted to this functional form using a least squares routine. The logarithmic terms are sometimes needed to obtain a good fit, but are weighted by $(1-z)$ noting that an expansion about $\gamma=0$ of equation (\ref{hypexpand}) is finite as $z\rightarrow 1$. This is also the case for the other hypergeometric functions encountered in the impact factors. \\

One can now expand in $N$ and $\gamma$ the $z$ integrands in equations (\ref{SLinttot}, \ref{STinttot}), after substituting in the parameterised hypergeometric functions. Then one has:
\begin{equation}
S_i(\gamma,N,M^2/Q^2)=\int_0^1dz\sum_n\left[S_{i,n}^{(0)}(z,M^2/Q^2)+NS_{i,n}^{(1)}(z,M^2/Q^2)\right]\gamma^n+{\cal O}(N^2)
\label{impactparam}
\end{equation}
where the power series in $N$ is truncated at NLL order. For given values of $M^2/Q^2$, the $z$ integrals can be calculated numerically to obtain the expanded massive impact factor at a given momentum scale. For example, choosing $M=1.5$GeV and $\Lambda=150$MeV for the QCD scale parameter, one finds at $t\equiv\log(Q^2/\Lambda^2)=7$, i.e. $Q^2=25$GeV$^2$:
\begin{align}
C_{Lg}(\gamma,N,M^2/Q^2)|_{t=7}&=\frac{\alpha_S}{4\pi}[.5184+.2069\gamma+.9468\gamma^2+.3799\gamma^3+1.155\gamma^4+N(-.4869-1.105\gamma-1.936\gamma^2\notag\\
&-2.880\gamma^3-3.785\gamma^4)];\label{powseriesL}\\
C_{2g}(\gamma,N,M^2/Q^2)|_{t=7}&=\frac{\alpha_S}{4\pi}[3.318+1.280\gamma+7.095\gamma^2+3.473\gamma^3+10.35\gamma^4+N(-5.154-8.613\gamma-20.22\gamma^2\notag\\
&-28.41\gamma^3-45.95\gamma^4)].
\label{powseries2}
\end{align}
One may check the LL coefficients in equations (\ref{powseriesL}, \ref{powseries2}) agree with the results of expanding the LL impact factor found in \cite{Catani2,CataniH}. Furthermore, one may verify the NLL part of $C_{Lg}$ for $M^2/Q^2\rightarrow 0$ against the massless exact kinematics result \cite{Peschanski}. One cannot do this for $C_{2,g}$ due to the fact that the transverse impact factor diverges as $M^2/Q^2\rightarrow0$. However, it is possible to ascertain the correct asymptotic behaviour. At LL order one has:
\begin{equation}
C_{2,g}^{LL}(\gamma,M^2/Q^2)|_{\frac{M^2}{Q^2}\rightarrow0}\longrightarrow f_1(\gamma)\left(\frac{M^2}{Q^2}\right)^\gamma+f_2(\gamma).
\label{asymptLL}
\end{equation}
At NLL level and beyond, one can make a similar ans\"{a}tz by promoting $f_i=f_i(\gamma,N)$. Lesser powers of $M^2/Q^2$ would vanish as $M^2/Q^2\rightarrow0$, and higher powers are inconsistent with the fact that the fixed flavour coefficient functions in $Q^2$ space contain at most collinear divergences $\sim\alpha_S^n\log^n(M^2/Q^2)$. From equation (\ref{asymptLL}) one finds:
\begin{align}
f_2(\gamma,N)&=\frac{1}{\gamma}\left[C_{2,g}(\gamma,N,M^2/Q^2)+\frac{\partial C_{2,g}(\gamma,N,M^2/Q^2)}{\partial\log{Q^2/M^2}}\right];\label{f2sol}\\
f_1(\gamma,N)&=\lim_{\frac{M^2}{Q^2}\rightarrow0}\left(\frac{M^2}{Q^2}\right)^{-\gamma}\left[C_{2,g}(\gamma,N,M^2/Q^2)-f_2(\gamma,N)\right].\label{f1sol}
\end{align}
There is also the consistency relation:
\begin{equation}
\left[\gamma \,C_{2,g}(\gamma,N,M^2/Q^2)+\frac{\partial C_{2,g}(\gamma,N,M^2/Q^2)}{\partial\log(Q^2/M^2)}\right]_{\frac{M^2}{Q^2}\rightarrow 0}\longrightarrow \gamma\,S'_2(\gamma,N),
\label{consist}
\end{equation}
where ${S'}_2(\gamma,N)$ is the massless impact factor with exact gluon kinematics. Thus one has $f_2(\gamma,N)={S'}_2(\gamma,N)$ as an all orders result in $\gamma$. Then $f_1$ can be found as a series expansion from equation (\ref{f1sol}) by substituting the power series for $C_{2,g}(\gamma,N,M^2/Q^2)$ at some sufficiently low value of $M^2/Q^2$. It can be verified against the result obtained in the LL case. Not only is this an important check of the calculation, but the asymptotic limit is of phenomenological importance for disentangling the heavy flavour coefficients in a variable flavour scheme at high $Q^2$ (see \cite{TW1} for how this works at LL order).\\

\begin{figure}
\begin{center}
\scalebox{0.8}{\includegraphics{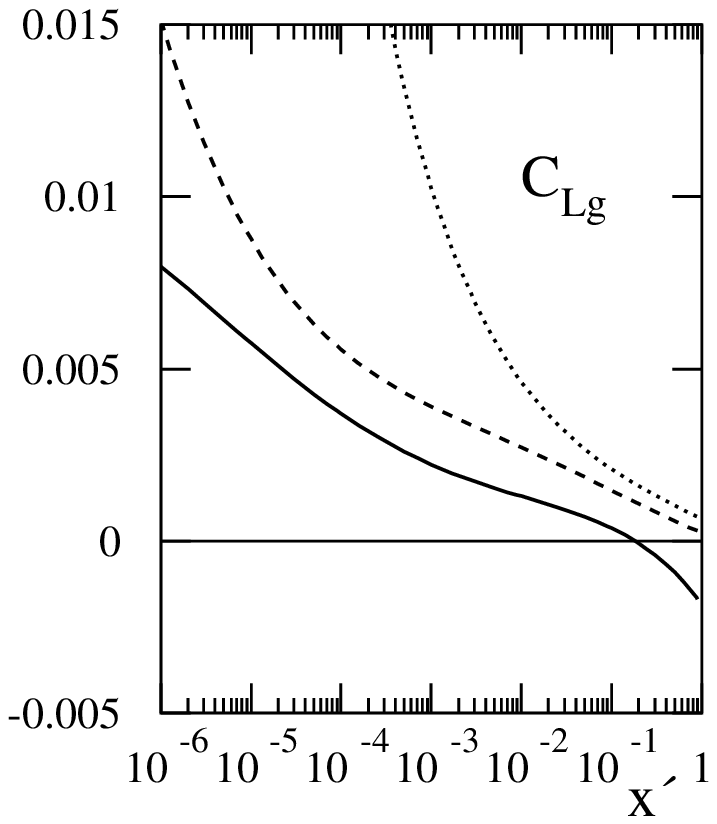}}
\scalebox{0.8}{\includegraphics{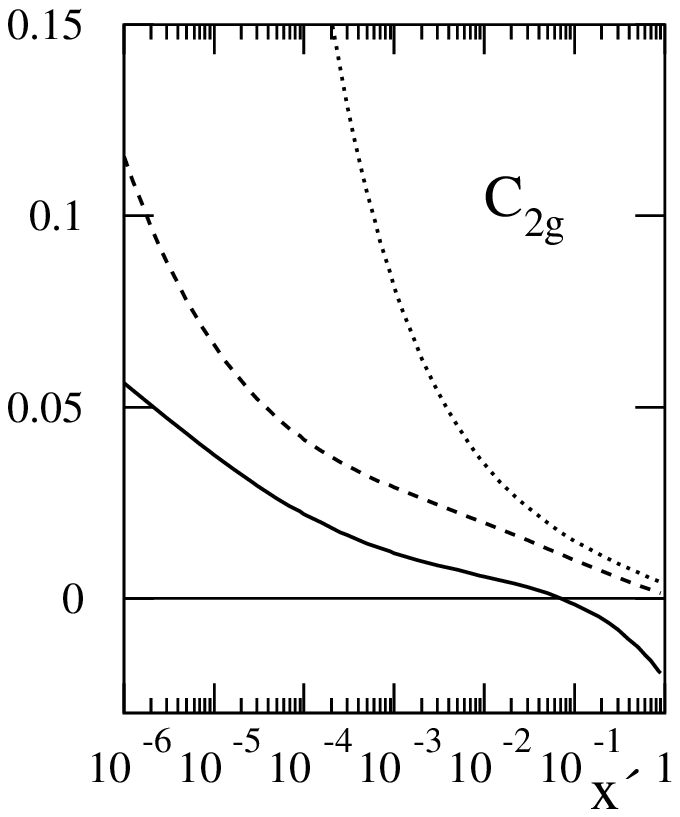}}
\caption{Estimates for the NLL resummed fixed flavour heavy gluon coefficient functions for $F_L$ and $F_2$ (solid) using the exact kinematic impact factors at $t=7$, $n_f=3$. Also shown is the LL result with running coupling corrections (dashed) and the strictly LL result (dotted).}
\end{center}
\label{plots}
\end{figure}

Armed with the series expansions of equations (\ref{powseriesL}, \ref{powseries2}), one can use the method of \cite{Thorne01} to obtain the corresponding estimated NLL coefficients $C_{i,g}$ in $x'$ and $Q^2$ space. Results are shown in figure \ref{plots} \footnote{There are also corrections $\sim{\cal O}(\Lambda^2/Q^2)$ to these results due to ambiguities in the derivation \cite{Thorne01}. We assume, however, that they are small at $t=7$.}. We also show the LL results with running coupling corrections and the strictly LL results. Note that all of these coefficients are in the {\it $Q_0$ scheme}, where the gluon is defined by the solution of the BFKL equation. In principle a transformation is needed to obtain the corresponding results in the more commonly used $\msbar$ scheme. Looking at the figure, one can see that the LL coefficients are strongly divergent at small $x$. This divergence is tempered significantly by the inclusion of the running coupling (as already noted in \cite{Thorne01,TW1}) and the effect of the approximate NLL corrections is to suppress the small $x$ divergence yet further. A similar suppression is observed in the massless case \cite{White}, and in \cite{TW1} it was noted that a softening of resummation effects in the moderate $x$ region is needed to achieve a good fit to scattering data.\\

We note that a smaller suppression of the small $x$ divergence occurs in the massive impact factor results than in the corresponding massless quantities when exact kinematics are included. One can examine this by comparing the factor $K_N^H$ of equation (\ref{KHN}) with the equivalent factor in the massless case, which is:
\begin{equation}
K_N^{M=0}=\frac{(\hat{Q}^2)^N}{[(\vec{p}-(1-z)\vec{k})^2+\hat{Q^2}+\hat{k}^2]^N}
\label{KN0}
\end{equation}
This has a maximum value of unity at $k^2=0$ and $(\vec{p}-(1-z)\vec{k})^2=0$. It then acts to suppress the integrand for the impact factor as $\vec{k}$ and $\vec{p}$ move away from these values, becoming noticeable at a scale given by $(\vec{p}-(1-z)\vec{k})^2+k^2\sim \hat{Q}^2$. Similarly, $K_N^H$ has a maximum value of unity at the same values of $\vec{k}$ and $\vec{p}$, and $z=1/2$. But the typical scale at which a significant suppression occurs is increased by the presence of the heavy quark mass to $(\vec{p}-(1-z)\vec{k})^2+k^2\sim \hat{Q}^2+4M^2$. Thus there is less of an effect in the massive case.\\

This reduction of the suppression due to exact kinematics in the heavy flavour structure functions 
is of particular phenomenological interest. NLO QCD fits have a tendency to undershoot data on $F_2^c$ at low $x$ (e.g. \cite{H1}). The LL resummed fit of \cite{TW1} also underestimated $F_2^c$ at very low $x$, and thus one would hope that in a NLL resummation the heavy coefficients are suppressed less with respect to the massless results. It is encouraging that this indeed seems to be the case. 
\section{Conclusions}
In this letter we have calculated the virtual photon-gluon impact factor with massive quarks and exact gluon kinematics, presented in equations (\ref{SLinttot}, \ref{STinttot}). Closed analytical results do not seem to be possible in $M^2/Q^2$ space (there is a remaining integration over the heavy quark momentum fraction $z$), although are possible in the scaled triple Mellin space of section \ref{tripmel} and are given in equations (\ref{SLtilde2}) and (\ref{STtilde2}).  Furthermore, one can still use the results in $M^2/Q^2$ space for phenomenology. As an example, we presented estimates for the NLL resummed fixed flavour heavy flavour gluon coefficients $C_{2,g}$ and $C_{L,g}$. The estimated NLL effects are seen to suppress the low $x$ divergence when compared with both the fixed and running coupling LL results. Such an effect is encouraging, as this is hopefully what is needed to achieve a good fit to structure function data over the complete $x$ range. Furthermore, the massive results are suppressed less than the corresponding massless results as a consequence of the kinematic constraint on the partonic centre of mass energy at the heavy quark vertex. This is consistent with previous global fits at NLO and LL orders in the QCD expansion that underestimate the charm data at very low $x$. Work on implementing our results in an approximate NLL global fit is ongoing. 
\section{Acknowledgements}
CDW is supported by a PPARC studentship and carried out part of this work in the SPhT at Saclay to whom he is extremely grateful for warm hospitality. RP wishes to thank A. Bialas and H. Navelet for fruitful collaboration. RST thanks the Royal Society for the award of a University Research Fellowship, and is also grateful to the HEP group at the Cavendish Laboratory for hospitality. 

\bibliography{refsp}
\end{document}